# Accessing the conduction band dispersion in CH$_3$NH$_3$PbI$_3$ single crystals

Jinpeng Yang [1, 2*], Haruki Sato[3], Hibiki Orio[3], Xianjie Liu[4], Mats Fahlman[4], Nobuo Ueno[5], Hiroyuki Yoshida[6,7], Takashi Yamada[8], Satoshi Kera[1*]

[1]Institute for Molecular Science, Department of Photo-Molecular Science, Myodaiji, Okazaki, Japan.

[2] College of Physical Science and Technology, Yangzhou University, Jiangsu, China.

[3] Graduate School of Science and Engineering, Chiba University, Chiba, Japan

[4]Laboratory for Organic Electronics, ITN, Linköping University, Norrköping, Sweden

[5]Graduate School of Advanced Integration Science, Chiba University, Chiba, Japan.

[6]Graduate School of Engineering, Chiba University, Chiba, Japan

[7]Molecular Chirality Research Center, Chiba University, Chiba, Japan

[8]Department of Chemistry, Graduate School of Science, Osaka University, Toyonaka, Japan

*Correspondence to: yangjp@yzu.edu.cn; kera@ims.ac.jp

**Abstract**

The conduction band structure in methylammonium lead iodide ($CH_3NH_3PbI_3$) was studied both by angle-resolved two-photon photoemission spectroscopy (AR-2PPE) with low-photon intensity and angle-resolved low-energy inverse photoelectron spectroscopy (AR-LEIPS). Clear energy dispersion of the conduction band along the ΓM direction was observed by these independent methods under different temperatures, and the dispersion was found to be consistent with band calculations under the cubic phase. The effective mass of the electrons at the Γ point was estimated to be $(0.20±0.05)m_0$ at 90 K. The observed energy position was largely different between the AR-LEIPS and AR-2PPE, demonstrating the electron correlation effects on the band structures. The present results also indicate that the surface structure in $CH_3NH_3PbI_3$ provides the cubic-dominated electronic property even at lower temperatures.

**Key words:** conduction band dispersion, angle-resolved low-energy inverse photoelectron spectroscopy, angle-resolved two-photon photoemission spectroscopy, organic halide perovskites

In the past few years, methylammonium lead iodide ($CH_3NH_3PbI_3$)-based organic-inorganic hybrid halide perovskite has emerged as one of the most promising photovoltaic absorbers due to its high-light absorption coefficient, long carrier lifetime and diffusion length, low exciton binding energy, and easy fabrication. [1-5] Compared with the development of device performance, the fundamental understanding in $CH_3NH_3PbI_3$ remains a subject of debate, [6-9] with several controversial points are still existing and waiting for urgent clarifications. In particular, experimental evidence for the occupied and unoccupied electronic states, which dominates device performances, is still not fully revealed. Very recently, Yang et al. [10] and Zu et al. [11] successfully reported the valence band (VB) structure for $CH_3NH_3PbI_3$ by using angle-resolved ultraviolet photoelectron spectroscopy (ARUPS), but the interpretation of the scanning direction in the Brillouin zone is controversial. In addition, temperature-induced gradual phase transitions have been confirmed either at the surface or in bulk, but there is an inconsistency of the observed top VB dispersion compared with surface structures. [10, 12] In order to further address these controversies, one effective way is to measure the electronic structures of the conduction band (CB), since the calculated band structure has directly indicated the different dispersive relations to the VB at various momentum spaces. To our best knowledge, no study on the CB structure of a $CH_3NH_3PbI_3$ single crystal exists so far because the thick single crystals commonly studied exhibit charging/damaging upon laser irradiation or electron bombardment [13, 14, 15]. For probing the electronic structure of the CB, angle-resolved two-photon photoemission spectroscopy (AR-2PPE) and angle-resolved inverse photoelectron spectroscopy (AR-IPES) are establised techniques; while AR-2PPE observes the conduction band with excitons, [16 17, 18], AR-IPES examines the conduction band with occupied anions. [19, 20] (More details can be found in Methods) However, the serious sample-damage and low energy resolution have prevented the CB measurements of $CH_3NH_3PbI_3$ single crystal using AR-IPES. These issues have been solved by lowering the electron and photon energies in low-energy inverse photoelectron spectroscopy (LEIPS). [21, 22] Recently, a new version of LEIPS having the angular resolution called angle-resolved low energy inverse photoelectron spectroscopy (AR-LEIPS) has been developed at Chiba University, which enables the CB measurement of $CH_3NH_3PbI_3$ single crystals.

In the present work, we have measured the CB dispersions of cleaved $CH_3NH_3PbI_3$ single crystals by either using low-photon fluxes for 2PPE or low-current density for LEIPS to overcome the sample charging and damaging. This has enabled us to accurately determine the dispersion relations along with defined crystallographic directions with different substrate temperatures, where the surface structure and the effective mass of electron can be quantitatively obtained. Moreover, the spectroscopic findings given by the different initial/final states demonstrate the impacts of charge relaxation processes on the electronic states, which offer to unveil a peculiar property of the hybrid perovskite materials, that is a long carrier lifetime.

**Fig. 1 (a)** depicts angle-integrated 2PPE spectra acquired at the normal-emission geometry with photon energy dependences measured at room temperature (300 K) to assign the conduction band related features in $CH_3NH_3PbI_3$ single crystal. The energies of these peaks (marked by bars in **Fig. 1(a)**) plotted as a function of photon energy are also given in **Fig. 1(b)**. Three different types of evolution in peak positions can be found in **Fig. 1(a)** (see also **Fig. S1** to describe transition processes of 2PPE). (i) Four peak positions (marked with A, $B_L$, $B_U$, and C) align vertically due to the two-step excitation, indicating these peaks arose from conduction bands, namely the energy positions above $E_F$ of 1.21 eV (denoted as $CB_0$), 1.82 eV (lower band: $B_L$ of $CB_1$), 1.98 eV (upper band: $B_U$ of $CB_1$), and 2.45 eV ($CB_2$), respectively. (ii) Peak D shifts gradually to a higher E-$E_F$ energy region with increasing photon energies. The shift changes linearly (as shown in **Fig. 1(b)**) with a slope of unity, indicating the peak is described from fixed valence band at 2.00 eV ($VB_0$) below $E_F$ via direct two-photon absorption, which is nearly consistent with UPS results (2.14 eV) by considering a different work function of the measured sample shown in **Fig. S2**. (iii) Peak E shifts oppositely to the lower E-$E_F$ energy region with increased photon energies. On the other hand, peak E can be well aligned once we put on the final energy of photoelectron with respect to $E_F$, as shown in **Fig. S3**, which suggests the origin of peak E is unoccupied states above the vacuum level. [23, 24] The work function measured from 2PPE is 4.48 eV (**Fig. S2**), which is nearly consistent with the result of 4.57 eV measured from UPS. Although no in-situ cleaning procedure in UHV was applied to the single crystal surface, our observations from 2PPE compared with results from UPS give us the direct information that the sample is of high quality and excellent surface cleanness. [10]

The conduction bands in the $CH_3NH_3PbI_3$ single crystal measured by AR-LEIPS at the normal incidence geometry for electrons with -5 V bias are also given in **Fig. 1**

**(c)**, where the beam damage and surface charging are confirmed to be negligible (as shown in **Fig. S4**). The spectrum is consistent with the previous LEIPS spectra of the solution-processed $CH_3NH_3PbI_3$ film. [25] The peak positions of spectral features of the $CH_3NH_3PbI_3$ single crystal are given after fitting with Gaussian functions, where the energy positions of 1.70 eV ($CB_0$), 2.34 eV ($CB_1$), and 2.89 eV ($CB_2$) above $E_F$ can be observed. **Fig. 1 (d)** summarizes the observed energy levels from 2PPE, LEIPS, and UPS (see **Fig. S2**). Importantly, the bottom energy position of the CB is different largely between the 2PPE (~1.2 eV) and LEIPS (~1.7 eV) to give the difference of about 0.5 eV [0.49 (±0.35) eV as described later]. Note that we do not discuss the shift of other CBs because of unclear spectral features in LEIPS and overlapping of the features via other transitions in 2PPE. The band-gap energy at the Γ point is estimated quantitatively by using angle-resolved data as described later. It is essential to notice that these methods involve different transition processes with electrons: 2PPE measurement includes both two-photon absorption as to give a similar binding energy to the UPS and two-step excitation processes from an initial state to a final state via an unoccupied intermediate state whereas LEIPS measurement reflects a process with materials generating from neutral (an initial state) to an anion (a final state). [20, 26] Interestingly, we here obtain nearly the consistent energy positions of $VB_0$ either from 2PPE or UPS due to the coherent two-photon absorption process in 2PPE with this photon energy range, in which the photoemission process is similar to that in UPS. On the other hand, $CB_0$ gives a clear energy difference of 0.49 (±0.35) eV by comparing these two methods (2PPE and LEIPS) due to the different initial/final state probing processes, where such energy difference could be considered as a half of on-site Coulomb energy if we assume in monitoring the band feature of upper Hubbard band in LEIPS. [20, 27]  However, the origin of the energy difference is not unambiguous for this class of materials where a charge couples strongly to phonons with a heavy carrier system, hence a general theoretical model should be extended. Importantly the observed band dispersions both VB and CB are different among these methods AR-2PPE, AR-LEIPS, and ARUPS in detail as described later. Note that the observed features in 2PPE could be ascribed from lowest exciton dispersion instead of conduction band dispersion upon forming a one-hole/electron paired final state, however, we conclude the dispersion character is evaluated for the conduction band because there is negligible difference in the energy position for the exciton band dispersion due to the extremely small exciton binding energy (0.02~ 0.05 eV) of the material [8, 28, 29], hence we use "CB" or "VB" labels for the related spectral features in the manuscript.

**Figs. 2 (a)** and **(b)** depict the photoelectron emission angle (θ) dependence of the AR-2PPE spectra of a $CH_3NH_3PbI_3$ single crystal. They are acquired separately at two different temperatures (300 K and 90 K) along the ΓM high symmetry direction (see the inset of **Fig. 3 (a)** for the measured direction in real space). Upon changing the photon energy at 90 K in 2PPE (see **Fig. S5**), the resonant photoemission from $VB_0$ to $CB_2$ is observed at around *hν* of 4.6 eV to give a clear separation of the VB and CB bands for the spectra taken at the larger *hν*. Therefore, the acquired photon energy is set to 4.77 eV for AR-2PPE measurements to resolve each peak in a spectrum. Each spectrum in **Figs. 2 (a)** and **(b)** is obtained with an acceptance angle of Δθ = ±1° to have a sufficient signal-to-noise ratio. Marked by a blue dash curve as a guide to the eye, a weak peak-like, energy-dispersive feature of the conduction band ($CB_0$) is visible in **Fig. 2 (a)**, where $CB_0$ is initially located at 1.36 eV ($\theta$= -8°), then reduces to 1.31 eV at normal emission ($\theta$= 0°), and starts to increases to 1.40 eV at $\theta$= 10°, with giving the total dispersion shift of ~0.1 eV. This dispersion of peak $CB_0$ can be more clearly recognized at a lower temperature (90 K) in **Fig. 2 (b)**. The $CB_0$ shifts to the low-energy (E-$E_F$) side overall and gives a larger dispersion than the 300 K results in **Fig. 2 (a)**, where $CB_0$ is initially located at 1.41 eV ($\theta$= -8°), then reduces to 1.09 eV ($\theta$= 0°) and again shifts back to 1.36 eV ($\theta$= 10°), with giving the total dispersion width of ~0.3 eV. On the other hand, no meaningful dispersion is found at both temperatures for the peaks $CB_1$. The dispersion relation of VB is differently observed between at 300 K and 90 K, namely no shift at 300 K, while a clear shift against the angle at 90 K. First, the $VB_0$ has a less dispersion in this momentum range as reported in the ARUPS taken at 300 K [10] (also see **Fig. S2**). Second, the feature is affected by overlapping the other CB features, like $CB_2$ bands (peak C), though the features could be distinguished by measuring the photon-energy dependence as shown in **Fig. 1** (300 K) and **Fig. S5** (90 K). Briefly, one can find some shoulder features around the peak $VB_0$ in the given photon energy of 4.74 - 4.77 eV. Increasing the peak width of $VB_0$ close to the Γ region also demonstrates the effect of other bands overlapping (see **Fig. S6**), hence we do not discuss changes in the dispersive features of $VB_0$ in detail. Different behavior in each peak shift of $CB_0$ and $CB_1$ is confirmed as shown in the bottom part of the figures as well. The spectral shift to the low-energy (E-$E_F$) side when reducing the temperature is also consistent with the previous ARUPS results for the VB change (seen **Fig. S2 (b)** and **(c)**), suggesting the work function change in $CH_3NH_3PbI_3$ upon the cooling. This temperature-induced peak shift ($VB_0$ and $CB_0$) indicates negligible impacts on the band-gap energy, as it may originate from an "n-type doping" behavior with pushing the Fermi level shift closer to the conduction band.

Incident angle ($\theta$) dependence of AR-LEIPS spectra along the ΓM high symmetry direction measured under two different temperatures (300 K and 170 K) have also been studied as shown in **Figs. 2 (c)** and **(d)**. Again a clear band dispersion for the conduction band ($CB_0$) is visible as shown in **Fig. 2 (c)**. The peak $CB_0$ is located at 1.81 eV ($\theta = -8°$), reaching the minimal position of 1.66 eV ($\theta = 0°$), and then shifts to a higher kinetic energy side with an increase of $\theta$, reaching 2.19 eV at $\theta=24°$. The dispersion becomes even more apparent when the measured temperature reduces to 170 K, as shown in **Fig. 2(d)**. Upon the cooling, the peak $CB_0$ is more prominent at $\theta= 0°$, which shows a larger dispersion than the 300 K with a minimal position of 1.25 eV and shifts to high-energy ($E-E_F$) side with the angles locating either at 1.45 eV ($\theta= -8°$) or 1.65 eV ($\theta= 32°$). The overall positions of $CB_0$ shift to the low-energy side upon the cooling due to work function change as found in 2PPE. The lower parts of **Fig. 2 (c)** and **(d)** demonstrate a direct comparison between two spectra recorded at two different angles ($\theta= 0°$ and $\theta= 20°$), where the dispersion of $CB_0$ is directly confirmed.

The bottom energy position of the concave CB, that is Γ point, is different significantly between the 2PPE (1.31 eV) and LEIPS (1.70 eV) at 300 K. The band-gap energy at the Γ point is estimated quantitatively. It is 3.84 eV from AR-UPS and AR-LEIPS which is larger than that of 3.23 eV obtained from AR-2PPE at 300 K. The inconsistency in the estimated-gap energy by various techniques has been discussed by considering the logarithm photoelectron spectra [30] and the band dispersion structure [11]. Accordingly, here we suggest the impact of electronic correlation on the band positions additionally. Theoretical development is highly encouraged to discuss the band renormalization. Note, the onset energy position has been considered to evaluate the band-gap energy in case of the sample has inhomogeneity in structure or consists of poly crystallites. Here we demonstrate in use the peak position instead of the onset position by momentum resolved data.

**Fig. 3** shows the second derivative map of the observed band dispersion along the ΓM direction plotted as a function of the parallel component of the electron wave vector $k_{//}$ measured by AR-2PPE at 90 K (a) and AR-LEIPS at 170 K (b). The calculated conduction band dispersion along the ΓM direction for a cubic phase (see **Fig. S7**) is also superimposed. The *E-k* dispersion maps taken at 300 K are shown in **Fig. S8**. A good agreement is obtained by comparing experimental results with the calculations in the conduction band region ($CB_0$ and $CB_1$). Moreover, combining with ARUPS results for VB dispersion (seen **Fig. S2**), the obtained CB's results strongly suggest the cubic-dominated surface structure in $CH_3NH_3PbI_3$ even at lower temperatures, which also agrees well with our previous work. [31] Note that we observe phase transition of the crystal structure with varying the temperature, where a tetragonal phase becomes visible below 250 K in a diffraction study [10], indicating the bulk structural properties are not dominant the surface electronic property. There is a controversy in the previous ARUPS reports [10, 11]; however, the concave shape of the band dispersion of $CB_0$ at the Γ point further confirms the measured dispersive direction in momentum space is strictly following the ΓM direction [10] instead of the XR direction [11], where a convex band ($CB_0$) should be found along the XR direction (see **Fig. S7**).

**Figs. 3 (c)** and **(d)** depict E-$k_{//}$ relation of the conduction band ($CB_0$) derived from AR-2PPE and AR-LEIPS results taken at lower temperatures (90 or 170 K). The calculated band dispersion ($CB_0$ with gray dash lines) is also superimposed in **Fig. 3 (c)**, in which a good agreement can be seen in these two spectra measured under different temperatures (see also **Fig. S8** for 300 K). However, in detail, the band dispersion is getting weaker around the Γ point at 300 K and deviates largely from the calculation at the high-k region at 90 K for 2PPE. The large difference of the band dispersion feature found in AR-2PPE compared to ARUPS and AR-LEIPS is caused by a peculiar property of hybrid perovskite materials, which is the strong electron-phonon coupling upon the exciton formation. [18] This is also supported why we can monitor 2PPE signals under the very low-laser power excitation, ten times weaker than a typical 2PPE experiment, which is realized by the nature of the long lifetime at the excited state. The deep insight into the lifetime is beyond the scope of this manuscript, and it is necessary to study time-dependent 2PPE experiments as feasible experiments in the future.

Furthermore, in **Fig. 3 (d)** we use a parabolic function mainly near the Γ point of the experimental curves to obtain the effective masses of electrons from the parabolic curvature according to the following equation: $m^*(k) = \hbar^2 \times [\partial^2 E(k)/\partial k^2]^{-1}$, where $m^*(k)$, $\hbar$, $E(k)$ and $k$ are the effective mass, reduced Planck's constant, energy, and wave vector, respectively. It is found that the effective electron mass $m_e^*$ to be 0.20 (±0.05) $m_0$ and 0.38 (±0.10) $m_0$ near the Γ point at 90 K by AR-2PPE and at 170 K by LEIPES, respectively, where $m_0$ is the free-electron mass. It is similar to be 0.28 (±0.10) $m_0$ and 0.39 (±0.10) $m_0$ for 2PPE and LEIPS at 300 K within the experimental error (**Fig. S8**). The larger $m_e^*$ of 0.38 (±0.10) $m_0$ in the LEIPS could suggest the observation of the upper Hubbard band upon forming a charged anion state, though we need further experiments as described before to demonstrate the band renormalization. As a feasible experiment in the future, it might be interesting to discuss the lower Hubbard band by having missing experiments on the time- and *hv*-dependent VB measurement with exciton formation in the 2PPE. Indeed the observed band shape in 2PPE assigned for $VB_0$ is largely changed upon the cooling to 90 K, and is different entirely from the theoretical calculation and the conventional ARUPS results.

In conclusion, we succeeded in observing the conduction band ($CB_0$) dispersion of $CH_3NH_3PbI_3$ single crystal near the Γ point utilizing AR-2PPE and AR-LEIPS for the first time, overcoming beam damage and surface charging. The energy difference of $CB_0$ between AR-2PPE and AR-LEIPS indicates a half of the on-site Coulomb energy, which is estimated to be 0.49 (±0.35) eV. Moreover, a clear conduction band dispersion, consistent with the calculations from a cubic dominated surface structure, has been confirmed at different temperatures. Our results should further inspire the extensive study of other types of hybrid perovskites to reveal conduction band structures related to orbital splitting [32] and phonon coupling effects.

**Methods**

$CH_3NH_3PbI_3$ single crystals (~2×1×10 $mm^3$) were grown with the method proposed by A. Poglitsch and D. Weber. [33] Afterward, the crystal structure and bulk quality were initially studied using crystal-X-ray diffraction (MERCURY CCD-1・R-AXIS IV), and the results were given in supplementary material (**SM**) **Table S1**. Measured samples were obtained by cleaving larger single crystals in an room atmosphere and put on a conductive silver paste adhered to a molybdenum sample holder [10], where we used the cleaved fresh surface for the measurements.

The UPS was performed using a high-sensitivity apparatus with a hemispherical electron energy analyzer (MBS A-1), monochromatic He $I_\alpha$ (hν = 21.22 eV) radiation source [10], where the resolution was set to 50 meV. A bias of -5 V was applied to the sample in order to detect the secondary cutoff electrons, whereas no bias for ARUPS measurements. The binding energy scale is referred to the Fermi level ($E_F$) measured on a clean metal substrate.

In AR-2PPE measurements with one-color irradiation for pump and probe photons, the p-polarized UV light was generated by frequency tripling the output of a titanium sapphire (Ti:Sa) laser operated at a repetition rate of 80 MHz and a pulse width of 110 fs. The resulting p-polarized light was focused on the sample surface with a concave mirror. The power of the laser light was set to ~0.0125 nJ/pulse at a repetition rate of 80 MHz, which was one order magnitude less than that used for conventional 2PPE measurements. [17] As a result, no damage to the sample and spectral changes due to the charging was found during the 2PPE measurements. The experiments was carried out in a home-built apparatus with an angle-resolved hemispherical energy analyzer (R3000, VG-Scienta, the emission acceptance angle in angle-resolved and angle-integrated mode is ±10° and 30°, respectively) and the overall energy resolution including the bandwidth of the laser light was typically less than 20 meV. [34] All 2PPE measurements were acquired with a sample bias of -3V to collect low-energy photoelectrons effectively. The influence of adding -3 V bias during AR-2PPE measurements has also been carefully considered, results in the underestimation of m* with a value of ~25% at most, especially when the value is evaluated far from the gamma point (see **Fig. S9**). [35, 36]

In AR-LEIPS measurement, an electron beam with the kinetic energy in the range between 0 and 5 eV from an electron source was incident to the sample surface. The emitted photons were collected and focused by a elliptical mirror into the photon detector consisting of a bandpass filter and a photomultiplier tube. The photon energy detected was 4.82 eV. The overall energy resolution measured from the Fermi edge of a polycrystalline Ag surface was 350 meV. No bias was added during AR-LEIPS measurements for dispersions. The beam damage in AR-LEIPS has been carefully avoided by reducing the emission current and slightly changing the measured sample position. The low-energy electron beam of low-current density (1~5 $\mu A/cm^2$) was irradiated on the sample to give a sample current of ~50 nA typically. Although we performed AR-LEIPS and 2PPE measurements for different sample crystals, we confirmed that these samples gave the same VB dispersions observed by separate experiments using ARUPS.

**Acknowledgements**

The authors thank Mr. K. Kawamura, Ms. H. Kuramochi of Chiba University for their assistance of the AR-LEIPS measurement. This work is financially supported in part by JSPS KAKENHI (No. JP26248062, JP18H03904, and JP26288007), and sponsored both by Qing-Lan Project from Yangzhou University and China Scholarship Council. A part of this work was performed with the aid of Instrument Center, Inst. Mol. Sci. Okazaki.

**Author contributions**

J. Y. and S. K. designed the experiments. J. Y. and T. Y. performed the experiments of AR-2PPE. J. Y., H. S., and H. O. performed the experiments of AR-LEIPS. J. Y. and X. L. performed the experiments of UPS and AR-UPS measurements. N. U., H. Y., T. Y. and M. F. contributed to content discussions. All the authors discussed the results and wrote the manuscript.

**Competing interests**

The authors declare no competing interests.

**Figures and captions**

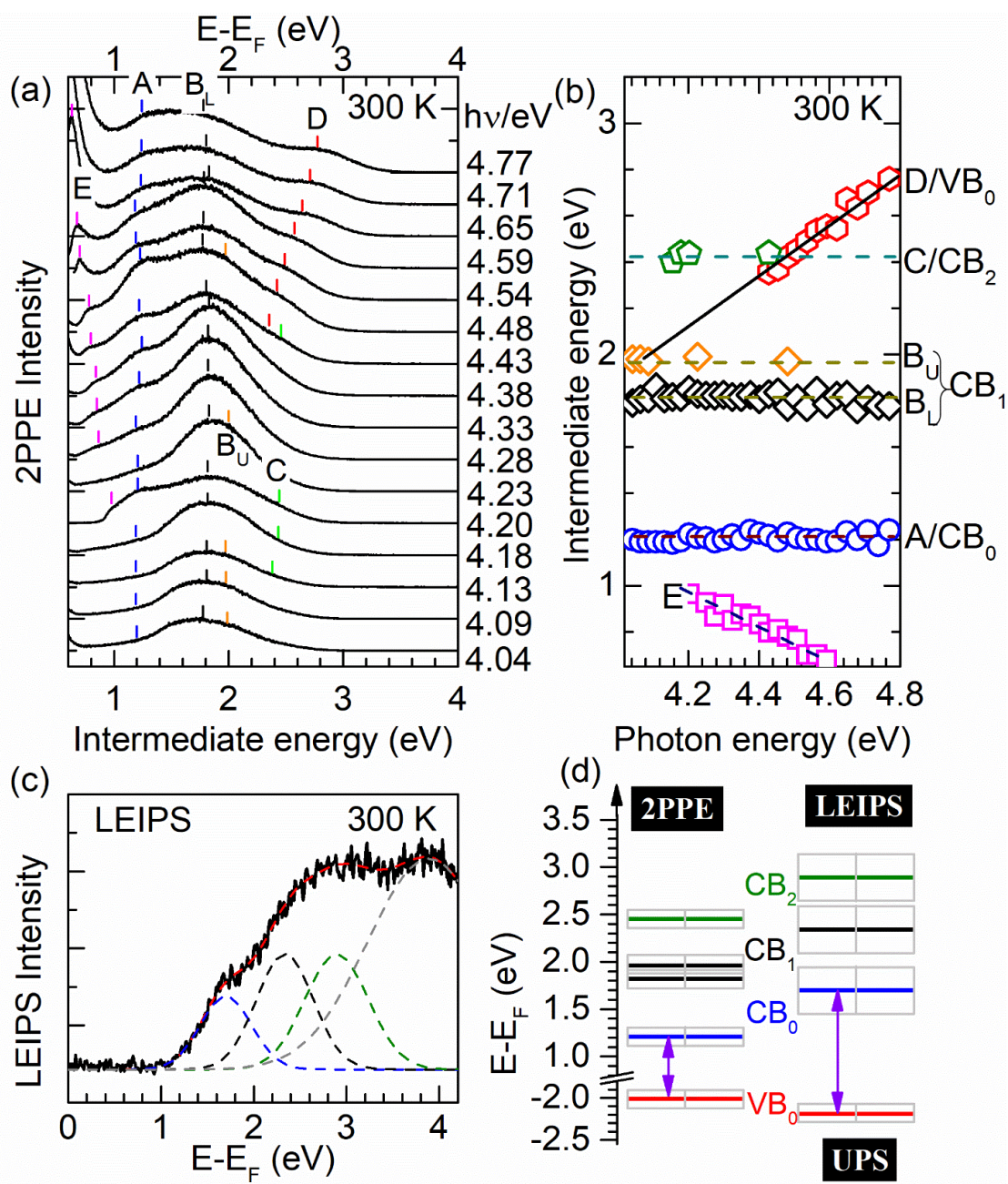

**Fig. 1** (a) Photon energy dependence of angle-integrated 2PPE spectra of CH$_3$NH$_3$PbI$_3$ single crystal measured under 300 K. The photon is irradiated for the incident angle of 60° with respect to the surface normal, and photoelectron is recorded at the normal emission geometry for photoelectrons with a sample bias of -3 V for acceptance angle of 30°. Photon energies are shown on the right-hand side. (b) Peak positions for 2PPE spectra features in (a) are plotted with respect to photon energy. Red hexagons lie on the line of slope 1Δ$h\nu$, and are assigned to the valence band (VB$_0$). Pink squares are assigned to final states, unoccupied states above the vacuum level (see **Fig. S2**). Peak positions of other spectroscopic features are independent of photon energies and are originated from the conduction band ($CB_0$, $CB_1$, and $CB_2$). (c) Angle-resolved LEIPS spectrum of $CH_3NH_3PbI_3$ at the normal incidence geometry for electrons using a bandpass filter of λ= 257 nm with a sample bias of -5 V. (d) Energy levels with peak position extracted from 2PPE (a), LEIPS (c), and UPS (see **Fig. S3**). The gray boxes are given with considering the energy and momentum resolutions of our measurements. The thick line indicates the mean value of the peak positions for various *hν* scans.

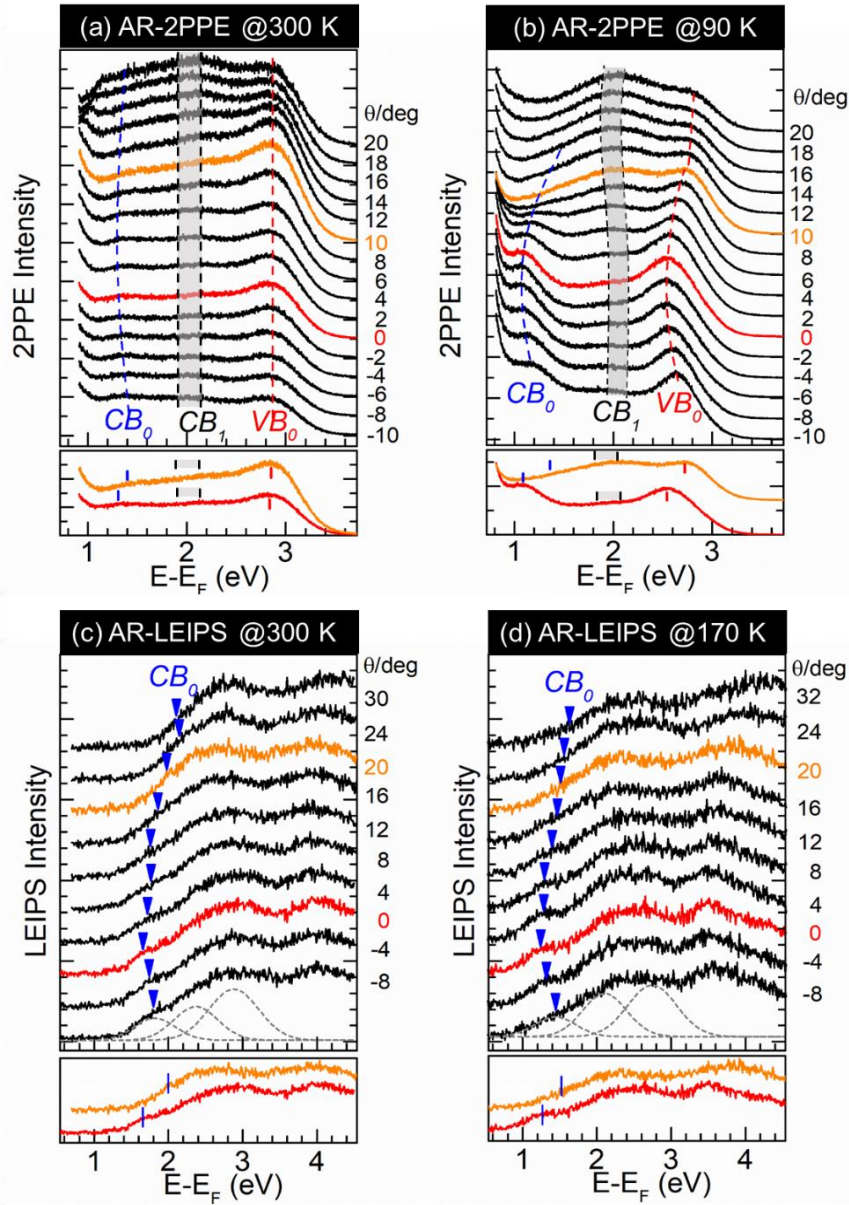

**Fig. 2.** AR-2PPE spectra (p-pol. *hv* of 4.77eV) of a $CH_3NH_3PbI_3$ single crystal along the ΓM direction with temperatures of 300 K (a) and 90 K (b) with a sample bias of -3 V (see **Fig. S9**). The band dispersion of each state ($CB_0$, $CB_1$, and $VB_0$) are marked with dash lines as guides to the eye. (c) and (d) depict the AR-LEIPS spectra of a $CH_3NH_3PbI_3$ single crystal along the ΓM direction with temperatures of 300 K and 170 K, respectively. No bias voltage was supplied during AR-LEIPS measurements. The gray dash lines are peaks fitted with Gaussian function, and peak positions ($CB_0$) at each θ is marked with a down arrow. The lower part of each figure gives a comparison between the spectra at θ=0° and θ=10° (or 20°) in order to clearly show the $E-E_F$ change due to the dispersion.

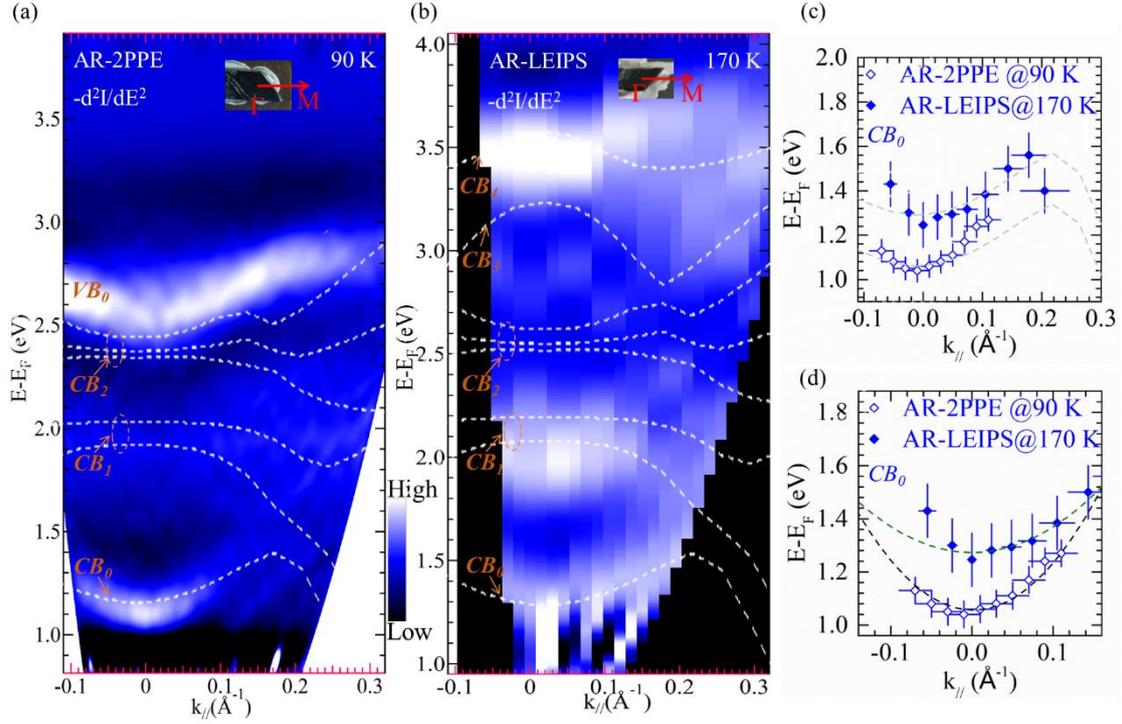

**Fig. 3.** (a) Second-derivative $E_k$-$k_{//}$ intensity maps of the AR-2PPE spectra (a) and the AR-LEIPS spectra (b) at low temperatures. Calculated dispersions for the conduction band ($CB_0$- $CB_4$) of a cubic phase are superimposed with dashed curves. (c) depicts E-$k_{//}$ relation of the conduction band ($CB_0$) derived from AR-2PPE and AR-LEIPS results taken at low temperatures (90 or 170 K), respectively. Theoretical curvatures ($CB_0$) obtained by the band calculation are also shown with gray dashed curves in (c). (d) The band curvatures ($CB_0$) close to the Γ point are estimated by parabolic-curve fitting (given by black and green dashed curves). The $CH_3NH_3PbI_3$ samples are shown on the insets of (a) and (b), respectively.

**Supplementary Material to**

**Accessing the conduction band dispersion in CH$_3$NH$_3$PbI$_3$ single crystals**

Jinpeng Yang [1,2*], Haruki Sato [3], Hibiki Orio [3], Xianjie Liu [4], Mats Fahlman [4], Nobuo Ueno [5], Hiroyuki Yoshida [6,7], Takashi Yamada [8], Satoshi Kera [1*]

[1]Institute for Molecular Science, Department of Photo-Molecular Science, Myodaiji, Okazaki, Japan.

[2] College of Physical Science and Technology, Yangzhou University, Jiangsu, China.

[3] Graduate School of Science and Engineering, Chiba University, Chiba, Japan

[4]Laboratory for Organic Electronics, ITN, Linköping University, Norrköping, Sweden

[5]Graduate School of Advanced Integration Science, Chiba University, Chiba, Japan.

[6]Graduate School of Engineering, Chiba University, Chiba, Japan

[7]Molecular Chirality Research Center, Chiba University, Chiba, Japan

[8]Department of Chemistry, Graduate School of Science, Osaka University, Toyonaka, Japan

**Supplementary material**

**Crystal X-ray diffraction measurement**

A small crystal of $CH_3NH_3PbI_3$ having approximate dimensions of ~0.200 × 0.200 × 0.200 mm was mounted on a glass fiber for single crystal X-ray characterization. The measurement was carried out on a Rigaku Mercury70 diffractometer using graphite monochromated Mo-Kα radiation at the Instrument Center of Institute for Molecular Science. The crystal-to-detector distance was 35.04 mm. The data was collected at a sample temperature of 350 ±0.5 K for $CH_3NH_3PbI_3$ to a maximum 2θ value of 59.9°, which was then analyzed by using the software CrystalClear (Rigaku). Fitting of the data using a cubic unit cell for $CH_3NH_3PbI_3$ yielded a total of 3564 reflections, in which 284 were independent (completeness =92.03%). The result was summarized in **Table S1**.

**Table S1.** Single crystal data for $CH_3NH_3PbI_3$

| Empirical Formula | $CH_6I_3NPb$ (350K) |
|---|---|
| Formula Weight | 619.98 |
| Crystal Dimensions | 0.200 × 0.200 × 0.200 mm |
| Crystal System | cubic |
| Lattice Type | Primitive |
| Lattice Parameters | a = 6.29 Å |
|  | V = 248.86 Å$^3$ |
| Space Group | P-43m (#215) |
| Z value | 1 |
| F000 | 260 |
| m(MoKα) | 261.937 cm$^{-1}$ |

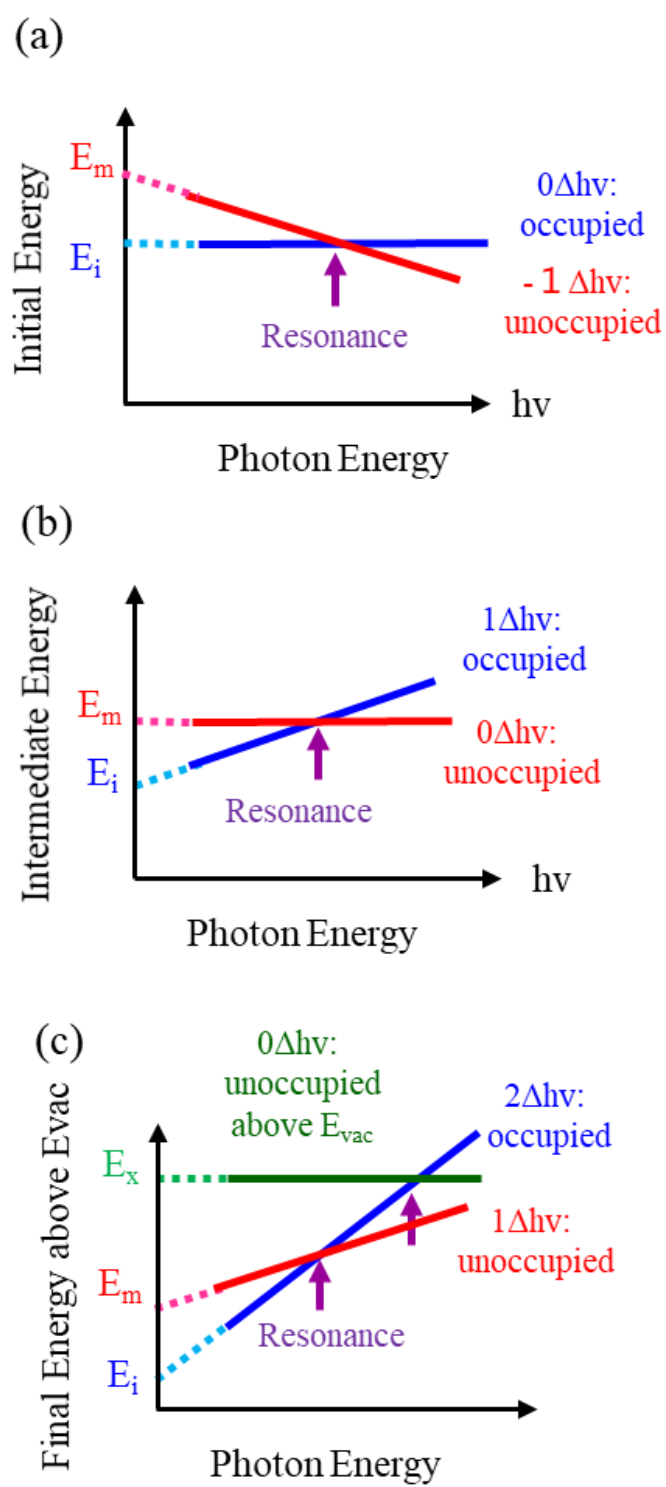

**Fig.S1** Energy diagram to define the observed state by considering the slope shape with different transition processes in 2PPE experiments.

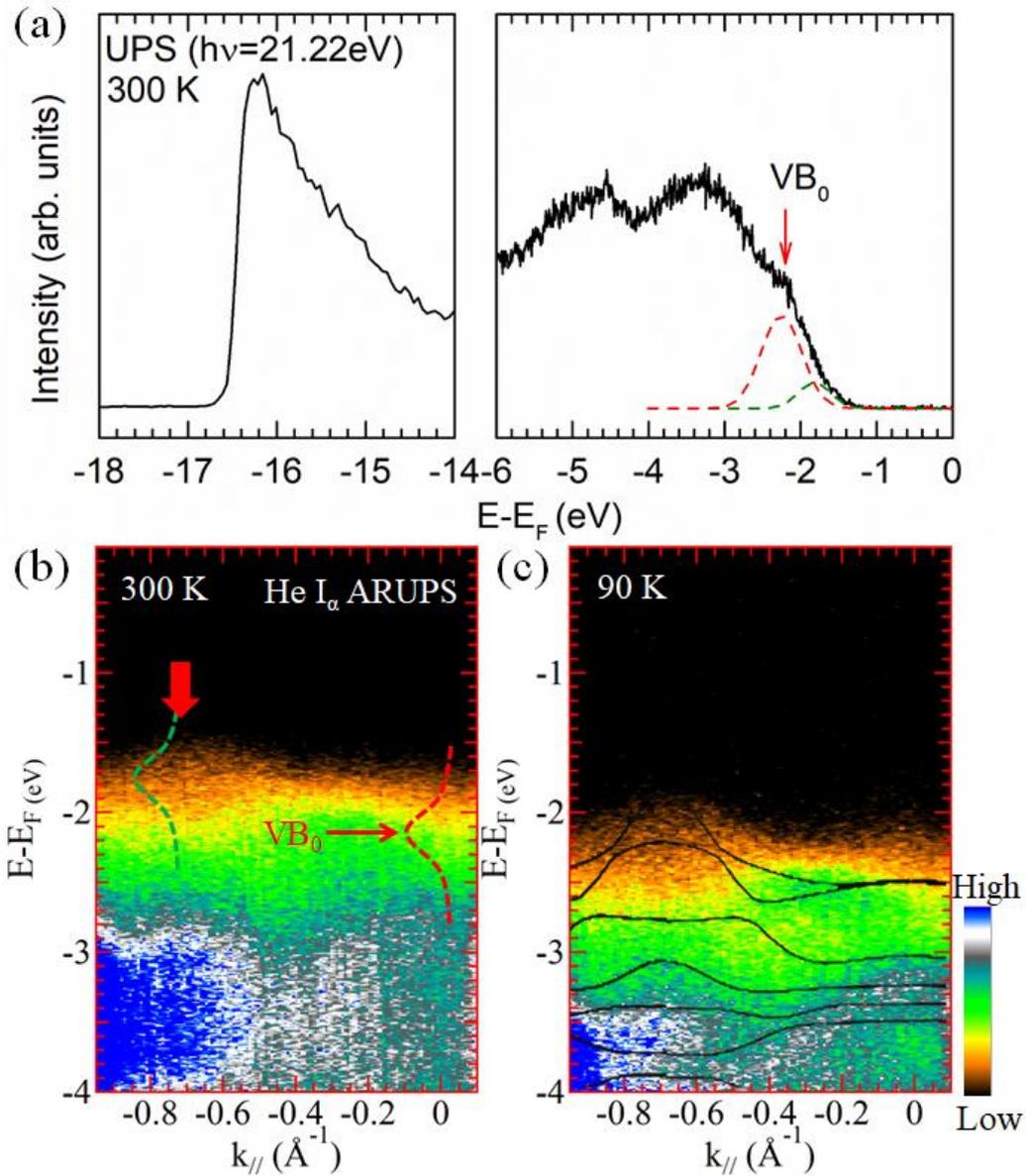

**Fig. S2.** (a) Angle-integrated He Iα-UPS spectrum of a $CH_3NH_3PbI_3$ single crystal measured at 300 K. The left side is on the secondary electron cutoff region and the right side on the valence band region. The red arrow indicates the peak position of the valence band ($VB_0$), which corresponds to the Γ point feature as found in (b). The position is nearly consistent with the 2PPE feature D in **Fig. 1.** (b) and (c) He Iα-ARUPS spectra of the $CH_3NH_3PbI_3$ single crystal along a cubic ΓM direction with temperatures of 300 K and 90 K, respectively. The red arrows indicate the M point (see also **Fig. S7**). Calculated dispersions for the valence bands of a cubic phase are superimposed with dashed curves in panel (c).

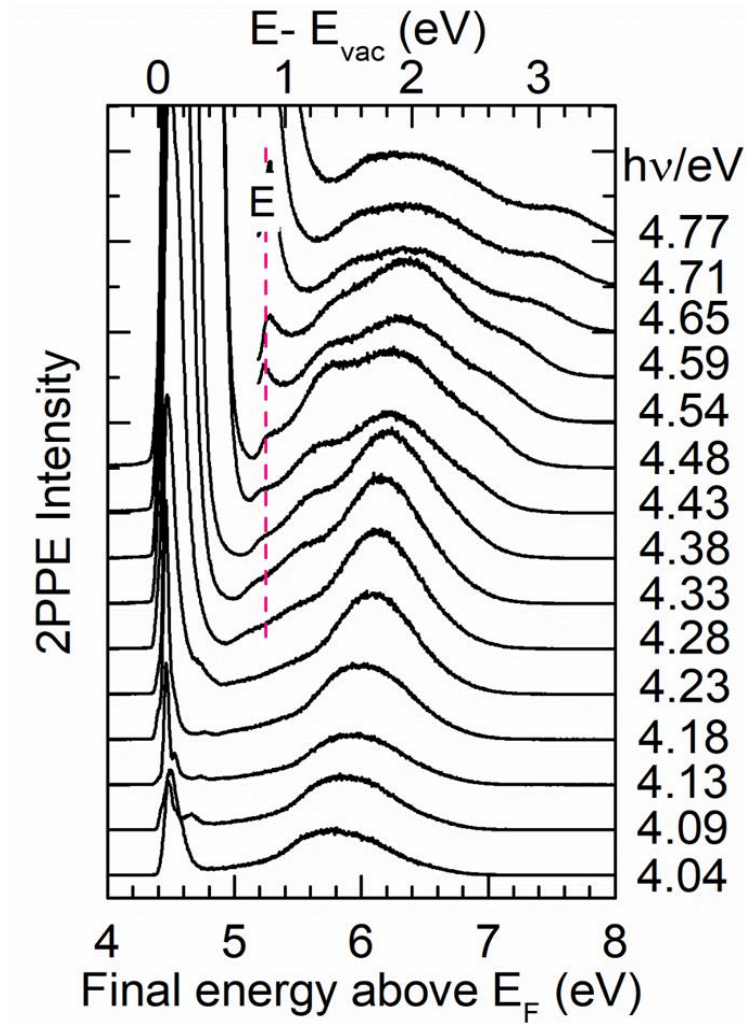

**Fig. S3.** Angle-integrated 2PPE spectra for a $CH_3NH_3PbI_3$ single crystal measured with *p*-polarized light. The photon energies are shown on the right-hand side. We can see peak E is well aligned with the final energy defined with respect to the $E_F$, and also aligned with the kinetic energy of photoelectron defined with respect to the vacuum level. This means the origin of the peak E is an unoccupied state located above the vacuum level (see also **Fig. S1**). The final energy above $E_F$ is defined as $(E-E_F)$ + (photon energy).

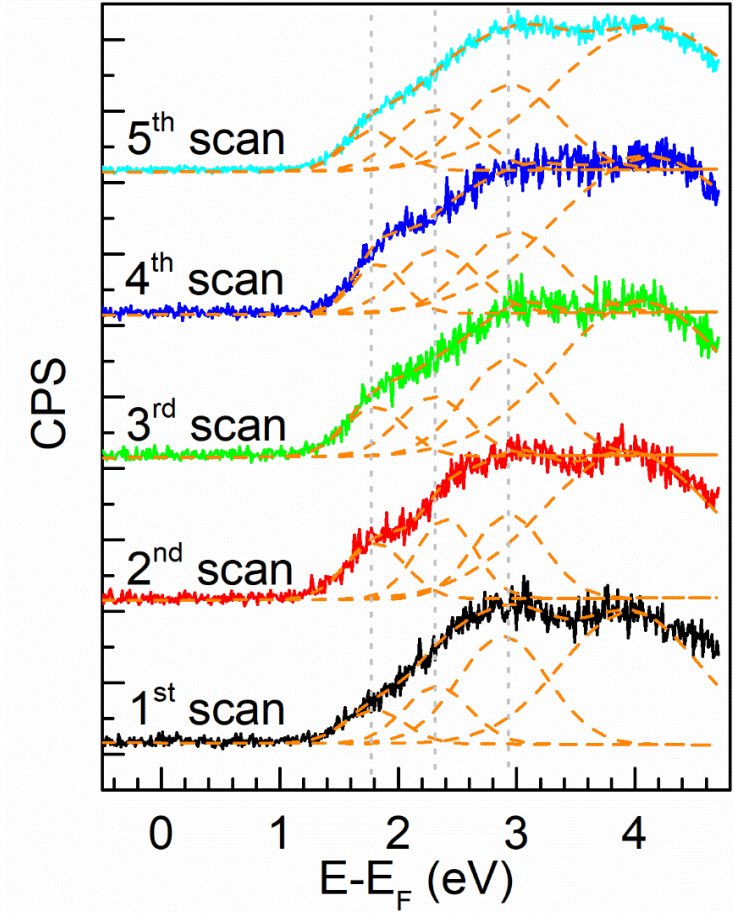

**Fig. S4.** LEIPS spectra measured for a $CH_3NH_3PbI_3$ single crystal with a series of scans. The sample position has not been changed during the measurement, where the negligible damage could be found after Gaussian fitting (Fitted parameters are given below in **Table S2**).

**Table S2.** Parameters used into fitting for **Fig. S4**.

|  | 1st scan | | | 2nd scan | | | 3rd scan | | | 4th scan | | | 5th scan | | |
| --- | --- | --- | --- | --- | --- | --- | --- | --- | --- | --- | --- | --- | --- | --- | --- |
|  | Peak 1 | Peak 2 | Peak 3 | Peak 1 | Peak 2 | Peak 3 | Peak 1 | Peak 2 | Peak 3 | Peak 1 | Peak 2 | Peak 3 | Peak 1 | Peak 2 | Peak 3 |
| Intensity (cps) | 23 | 40 | 75 | 39 | 55 | 58 | 34 | 41 | 67 | 34 | 44 | 56 | 27 | 42 | 59 |
| Position (eV) | 1.8 | 2.3 | 2.9 | 1.8 | 2.4 | 2.9 | 1.8 | 2.3 | 2.9 | 1.8 | 2.4 | 2.9 | 1.8 | 2.3 | 2.9 |
| Width (eV) | 0.60 | 0.66 | 0.89 | 0.60 | 0.61 | 0.73 | 0.60 | 0.69 | 0.89 | 0.59 | 0.72 | 0.91 | 0.58 | 0.75 | 0.93 |

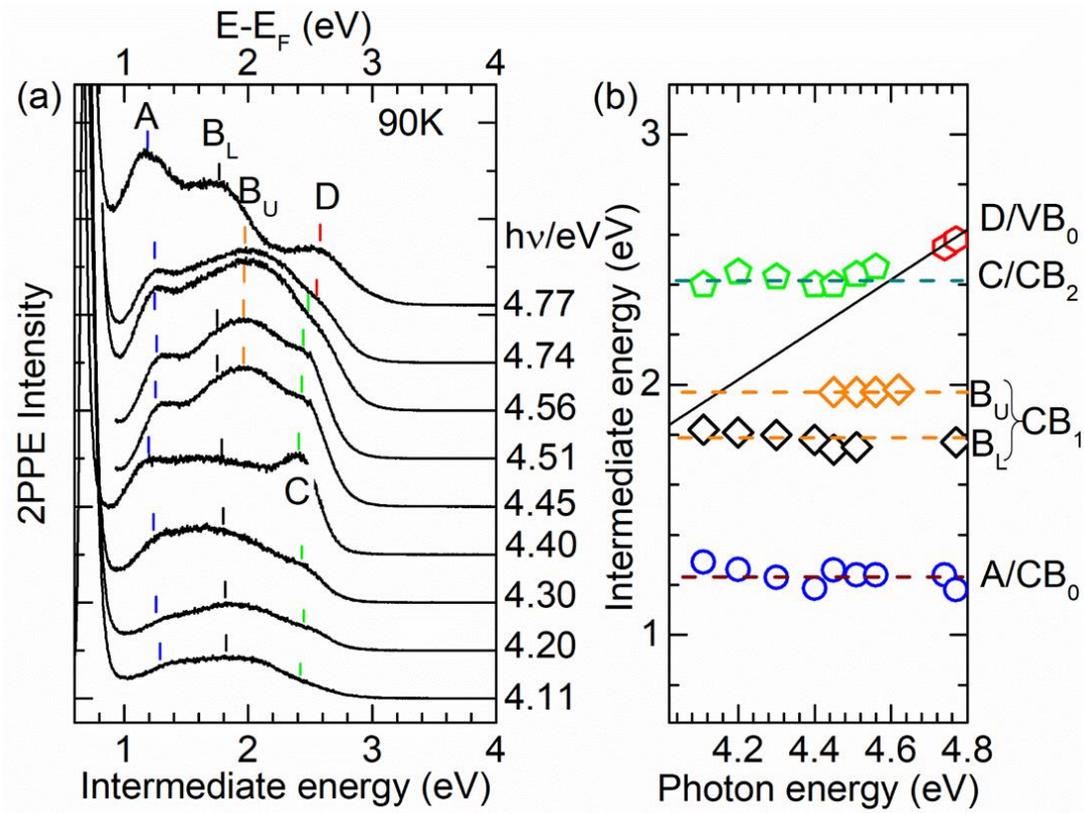

**Fig. S5.** Photon-energy dependence of angle-integrated 2PPE spectra of $CH_3NH_3PbI_3$ single crystal measured under 90 K. Photon energies are shown on the right-hand side. (b) Peak positions for 2PPE spectra features in (a) are plotted with respect to photon energy. These results at 90 K are consistent with those at 300 K in **Fig 1**.

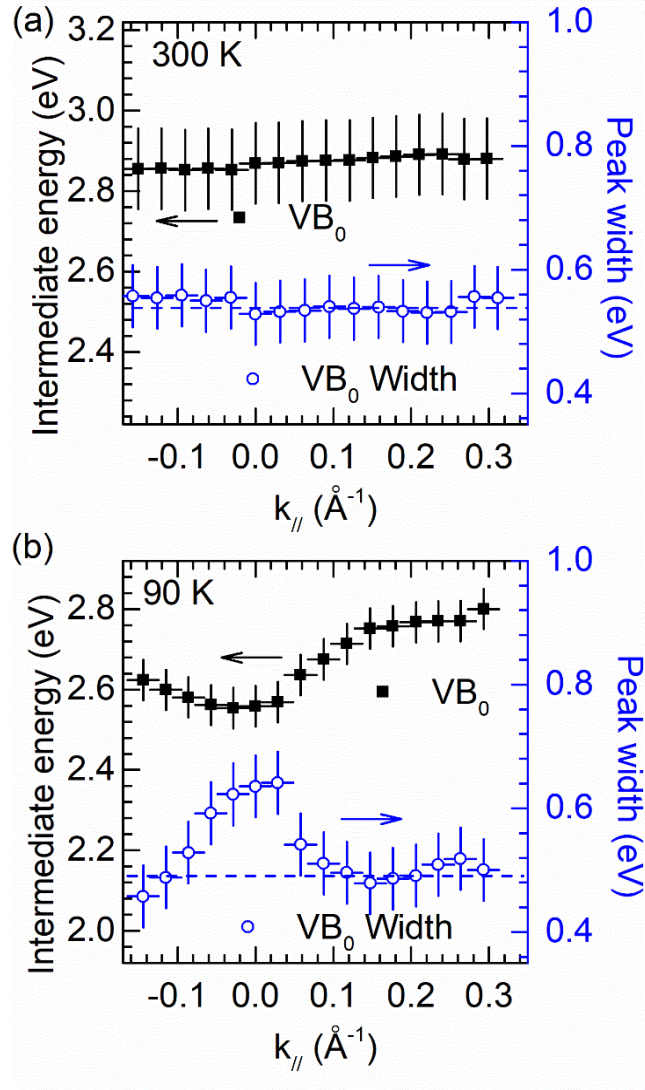

**Fig. S6.** (a) The valence band (VB$_0$) peak position (left side) and peak width (right side) after fitting from AR-2PPE spectra measured at 300 K; (b) VB$_0$ peak position (left side) and peak width (right side) extracted from AR-2PPE spectra measured at 90 K.

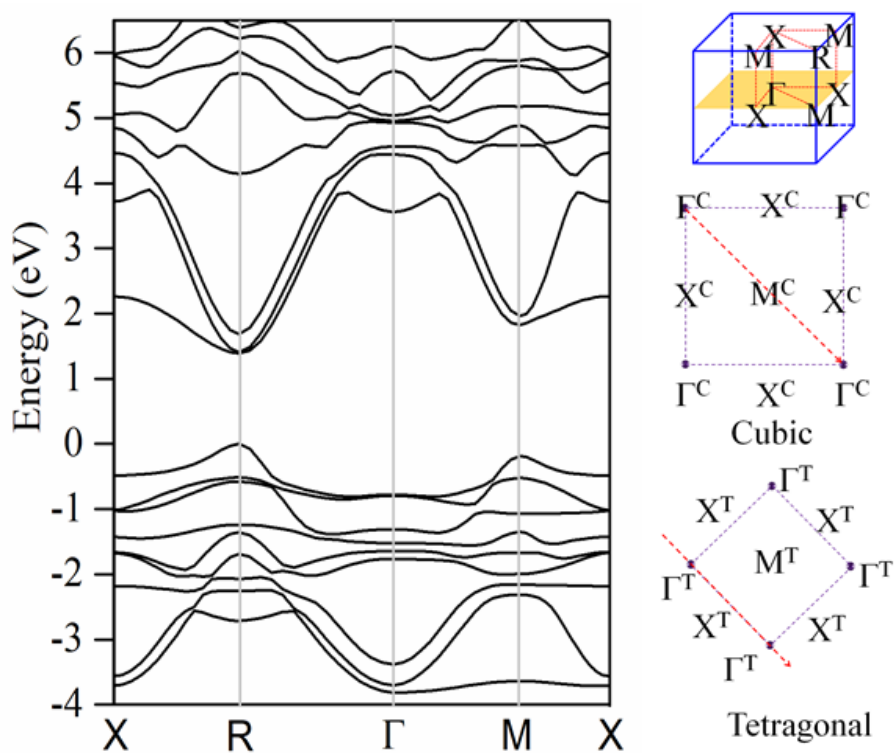

**Fig. S7.** On the left side: theoretical calculations of $CH_3NH_3PbI_3$ with a cubic phase. The density functional theory calculations were first performed with Perdew–Burke–Ernzerhof (PBE) generalized gradient approximation (GGA) as implemented in the Vienna Ab Initio Simulation Package (VASP). To obtain the energy gap, the screened hybrid functional method in HSE06 level was used. An energy cutoff of 600 eV, 12×12×12 and 4×4×4 Monkhorst-Pack grids were used in the PBE calculations. Schematic view of a crystal axis of scanning direction and surface BZs are shown. On the right side: the superscripts of T and C mean for tetragonal and cubic phase, respectively.

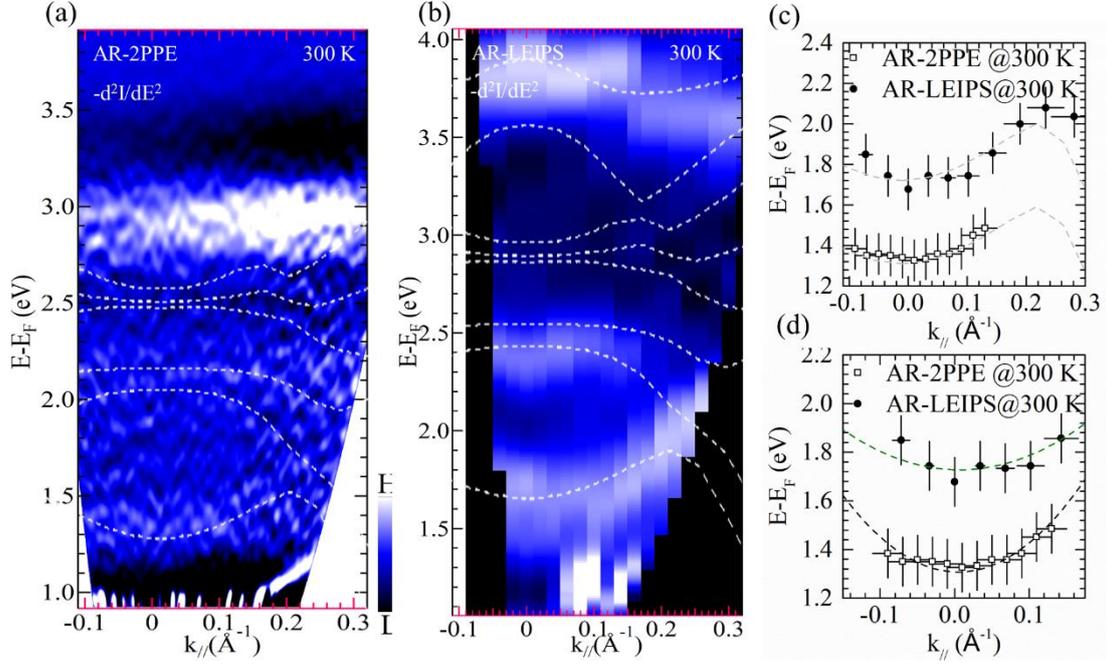

**Fig. S8** Second-derivative $E_k$-$k_{//}$ intensity maps of the AR-2PPE (a) and the AR-LEIPS (b) spectra at room temperature. Calculated dispersions for the conduction band ($CB_0$, $CB_1$, and $CB_2$) of a cubic phase are superimposed with dashed curves. (c) depicts E-$k_{//}$ relation of the conduction band ($CB_0$) derived from AR-2PPE and AR-LEIPS results taken at room temperature (300 K). Theoretical curvatures ($CB_0$) obtained by the band calculation are also shown with gray dashed curves in (c). (d) The band curvatures ($CB_0$) close to the Γ point are estimated by parabolic-curve fitting (given by black and green dashed curves). The m* is evaluated 0.28 (±0.10) $m_0$ and 0.39 (±0.10) $m_0$ for 2PPE and LEIPS at 300 K within the experimental error.

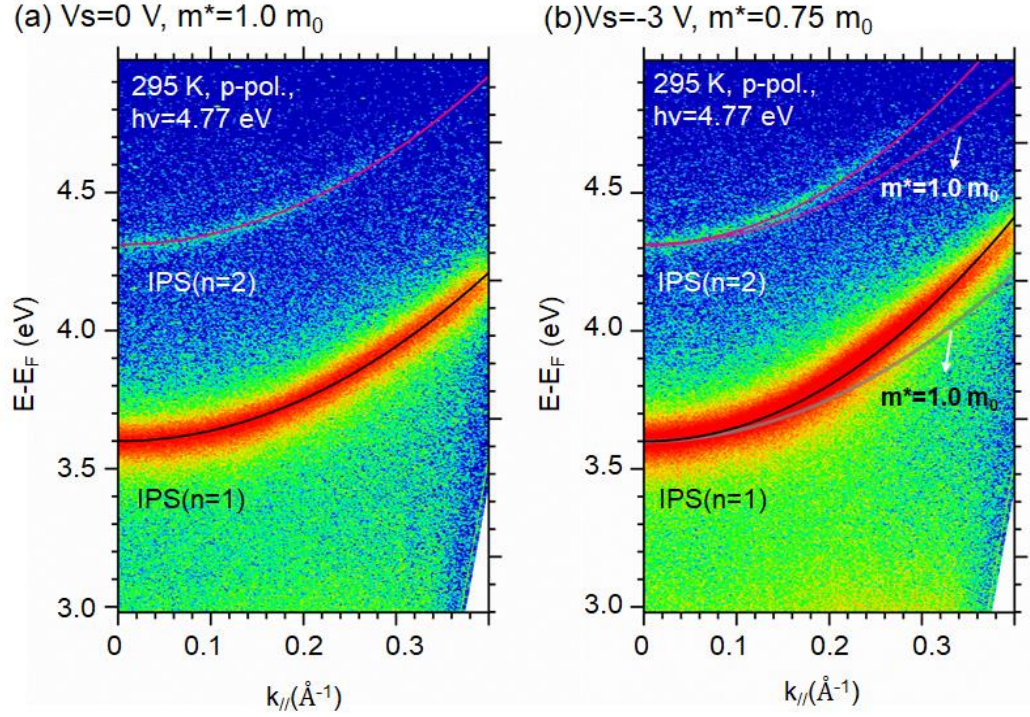

**Fig. S9.** Angle-resolved 2PPE maps of highly oriented pyrolytic graphite (HOPG) are shown measured (a) without sample bias (0 V), and (b) with sample bias (-3 V), respectively. The first (n=1) and second (n=2) members of image potential states (IPS) were observed. Without additional electric field in (a), IPS (n=1 and 2) showed nearly free-electron-like momentum-energy dispersions with effective masses of $m^*/m_0$ of ~ 1.0, where $m_0$ indicates the electron rest mass. The results were in agreement with previous studies [1-4]. Under the case with a sample bias of -3V in (b), the effective mass values were underestimated about 25% at most. The deviation is larger when the value is evaluated away from the gamma point (, i.e., $|k_{//}| > 0.2$ Å$^{-1}$). In this study, the effective mass values are evaluated in the region of $|k_{//}| < 0.15$ Å$^{-1}$, and thus we concluded that the estimated maximum deviations are compiled in the experimental errors shown in the main text.